\documentclass[twocolumn,showpacs,superscriptaddress,pra]{revtex4}
\usepackage{bm,graphicx,amsmath, subfigure}
\usepackage{ulem} 

\begin{document}

\title{The quantum ground state of self-organized atomic crystals in optical resonators}

\author{Sonia Fern\'{a}ndez-Vidal}
\affiliation{Grup d'\`{O}ptica, Departament de
F\'{i}sica, Universitat Aut\`{o}noma de Barcelona, E-08193
Bellaterra, Spain}

\author{Gabriele De Chiara}
\affiliation{Grup d'\`{O}ptica, Departament de
F\'{i}sica, Universitat Aut\`{o}noma de Barcelona, E-08193
Bellaterra, Spain}
\affiliation{Grup de F\'isica Te\`orica, Departament de
F\'{i}sica, Universitat Aut\`{o}noma de Barcelona, E-08193
Bellaterra, Spain}

\author{Jonas Larson}
\affiliation{NORDITA, 106 91 Stockholm, Sweden}

\author{Giovanna Morigi}
\affiliation{Grup d'\`{O}ptica, Departament de
F\'{i}sica, Universitat Aut\`{o}noma de Barcelona, E-08193
Bellaterra, Spain}
\affiliation{Theoretische Physik, Universit\"at des Saarlandes, D-66041 Saarbr\"ucken, Germany}

\date{\today}

\begin{abstract}
Cold atoms, driven by a laser and simultaneously coupled to the quantum field of an optical resonator, may self-organize in periodic structures. These structures are supported by the optical lattice, which emerges from the laser light they scatter into the cavity mode, and form when the laser intensity exceeds a threshold value. We study theoretically the quantum ground state of these structures above the pump threshold of self-organization, by mapping the atomic dynamics of the self-organized crystal to a Bose-Hubbard model. We find that the quantum ground state of the self-organized structure can be the one of a Mott-insulator or a superfluid, depending on the pump strength of the driving laser. For very large pump strengths, where the intracavity intensity is maximum and one would expect a Mott-insulator state, we find intervals of parameters where the system is superfluid. These states could be realized in existing experimental setups.
\end{abstract}

\maketitle

\section{Introduction}

Self-organization in systems of atoms and light has been observed in numerous experiments, some of which are reported in Refs.~\cite{Kaiser,OptLatt,Bistability,CARL,Black_03,Esslinger_09}. Several experimental realizations show the formation of spatially-ordered atomic structures, which organize in the potential they form~\cite{OptLatt,Black_03,Esslinger_09}. The basic mechanism behind the observed dynamics can be summarized by considering that the refractive index of the atomic medium, which is related to the atomic spatial density, is itself determined by the light fields via the mechanical effects of atom-photon interactions. Hence, the effective dynamics the atomic center of mass undergoes is determined by potentials and/or forces, which in return depend on the atom position and velocity distributions~\cite{Domokos_JOSAB03}.

In this context, one remarkable example is the formation of regular patterns of atoms in the standing wave of a high-finesse optical cavity, arising when the atoms are transversally driven by lasers. Experimental signatures are the phase-locking between the field at the cavity output and the driving laser phase, according to two possible values with difference equal to $\pi$. This occurs when the laser frequency is smaller than the cavity-mode frequency and its intensity exceeds a threshold value~\cite{Black_03}. The phenomenon, first predicted by numerical simulations according to a semiclassical model for atoms and light~\cite{Domokos_PRL02}, can be understood in terms of the atoms being trapped in the potential which originates from the light coherently scattered from the laser into the cavity mode. The largest intracavity amplitude is supported by the atomic configurations in which all atoms scatter in phase into the resonator, corresponding to the patterns where the interatomic distance is a multiple of the wavelength of the cavity field. Since the possible configurations which fulfill this condition in a standing-wave cavity are two, one shifted with respect to the other by half a wavelength, the difference between the possible phases of the emitted field is exactly $\pi$. A further theoretical study discussed the phenomenon in terms of second-order phase transition and determined the pump threshold within a semiclassical, mean-field model~\cite{Asboth_PRA05}.

Additional novel phenomena arise in this system when the quantum mechanical properties of light and matter are relevant. Theoretical works studied the dynamics of entanglement between atoms and fields during self-organization in a similar setup~\cite{Vukics_NJP07,Maschler_OC07,Maschler_EPJD08}. In Ref.~\cite{Domokos_EPJD08} the pump threshold for self-organization at $T=0$ was determined when the atoms are assumed to be forming a Bose-Einstein condensate. Most recently, the onset of self-organization has been experimentally observed in a Bose-Einstein condensate coupled to an optical resonator~\cite{Esslinger_09}. In such situation, an open question regards the nature of the quantum state of the system when the pump intensity is increased well above the self-organization threshold value. In this regime, in fact, the height of the potential increases, and one would expect localization of the atoms at the potential minima when the atom number is increased and hence the field in the cavity is larger. On the other hand, interparticle interactions due to $s$-wave collisions compete with such dynamics, favoring situations in which the number of atoms per site is small. This framework is reminiscent of the paradigmatic Mott-insulator / superfluid quantum phase transition~\cite{Fisher_PRB89}, and in particular its realization with ultracold atoms in optical lattices~\cite{Zwerger_RMP08}. In the case here considered, however, the dependence of the potential on the atomic distribution (and in particular, the expectation that its depth increases with the atomic density) makes it {\it a priori} unclear whether incompressible states can exist at large laser intensities.

\begin{figure}[htbp]
\begin{center}
\includegraphics[scale=0.55]{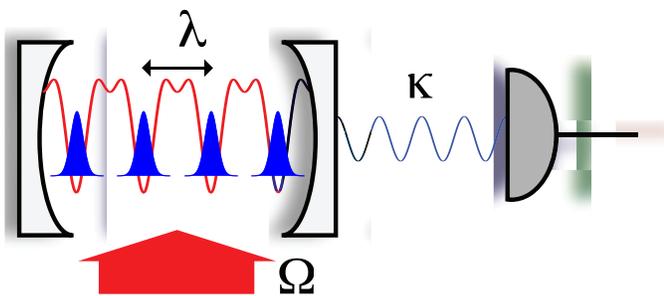}
\caption{(color online) Setup of the system, showing matter-wave density distribution (blue region) in the effective potential created inside an optical resonator by coherent scattering of laser photons (red line). Here, $\Omega$ is the Rabi frequency, giving the strength of the coupling between the atoms and the laser field, $\lambda$ is the wavelength of laser and cavity mode, and $\kappa$ is the rate at which photons leak from the cavity mode. The quantum state of the system could be inferred by measuring the light at the cavity output, or by Bragg spectroscopy using a weak probe (see Sec.~\ref{Conclusions}).}
\label{Fig:1}
\end{center}
\end{figure}

In this article we study the quantum ground state of this self-organized system, in the setup sketched in Fig.~\ref{Fig:1}. In particular, we determine under which conditions incompressible states exist at large pump intensities. For this purpose, we extend a procedure, developed for an atom-cavity system in which the cavity mode is pumped by a laser~\cite{Maschler_PRL05,Larson_PRL08,Larson_NJP08}, and derive a Bose-Hubbard model for the atoms trapped in the potential their generate, whose coefficients depend on the atomic density. Using the strong-coupling expansion~\cite{Monien_94}, we identify the parameter region where incompressible states of the self-organized atomic pattern may exist. This study finds particular motivation from experimental progress, which achieved the coupling of ultracold atoms with the optical mode of high-finesse cavities~\cite{Colombe_Nature07,Slama_PRA07,Stamper-Kurn_PRL07,Stamper-Kurn_NP08,Esslinger_Nature08,Esslinger_Science08,Esslinger_09}. Indeed, we will argue that incompressible self-organized states could be observed in the setups of existing experiments~\cite{Ritter_APB09,Esslinger_09}.

This article is organized as follows. The equations at the basis of the physical model are introduced in Sec.~\ref{model}, and the relevant approximations leading to the derivation of the effective atomic dynamics in the light potential are extensively discussed. In Sec.~\ref{Sec:Bose-Hubbard}, we assume that the atoms are tightly bound at the minima of the cavity potential inside the cavity, and derive an effective Bose-Hubbard model, whose coefficients are analytically determined within a modified Gaussian ansatz. The parameter regimes, where incompressible states of the self-organized atomic gas are found, are derived in Sec.~\ref{Results} within the strong coupling expansion method~\cite{Monien_94}. Conclusions and outlook are presented in Sec.~\ref{Conclusions}, while the appendix provides further details complementing the discussion in Secs.~\ref{Sec:Bose-Hubbard} and~\ref{Results}.

\section{The model}\label{model}

Identical bosonic atoms, with mass $m$, are confined inside an optical resonator, which is assumed to be a high-finesse cavity sustaining well resolved modes. The atoms are prepared at ultralow temperature $T$ and interact with each other by means of $s$-wave scattering. Furthermore, the atoms couple to light via a dipolar transition between the electronic ground and excited state $|g\rangle$ and $|e\rangle$ at the optical frequency $\omega_0$. In particular, the atomic dipole is simultaneously driven by a laser and by the field of one cavity mode, both linearly polarized and whose wave vectors are perpendicular to one another, according to the geometry of the setup sketched in Fig.~\ref{Fig:1}. We assume the atoms to be strongly confined in the plane perpendicular to the cavity axis, here chosen to be the $x$-axis, so that the motion is essentially along $x$. In absence of the resonator, the atomic dynamics for the atoms in states $|g\rangle$ and $|e\rangle$ are governed by the Hamiltonian ${\cal H}_0+{\cal H}_{eg}$, where
\begin{eqnarray}
\label{H:0}
{\cal H}_0&=&\sum_{j=g,e}\int\!dx\,\\&&\Psi_j^{\dag}\left(x\right)\left[-\frac{\hbar^2 }{2m}\frac{\partial^2}{\partial x^2}+\frac{g_{j}}{2}\Psi_j^{\dag}\left(x\right)\Psi_j\left(x\right)\right]\Psi_j\left(x\right)\,,\nonumber
\end{eqnarray}
with $g_{j}$ the strength of the state-dependent collisional interaction, and where ${\cal H}_{eg}$ describes $s$-wave scattering between atoms in different atomic states. The Hamiltonian in Eq.~\eqref{H:0} is written in second quantization, where operators $\Psi_i\left(x\right)$ and $\Psi_i^{\dag}\left(x\right)$ annihilate and create an atom at the position $x$ and in the electronic state $i=g,e$. The atomic field operators obey the bosonic commutation relations $\left[\Psi_i(x),\Psi^{\dag}_j(x')\right]=\delta_{ij}\delta\left(x-x'\right)$ and $\left[\Psi_i(x),\Psi_j(x')\right]=0$.

The cavity is assumed to be a high-finesse resonator sustaining well-resolved modes, of which the mode at frequency $\omega_c$ is (quasi) resonant with the atomic transition. We denote by  $a$ and $a^{\dagger}$ the annihilation and creation operators of a cavity photon at frequency $\omega_c$. The coherent coupling between atoms and light is described by a Hamiltonian in the Rotating-Wave approximation, which reads
\begin{eqnarray}
\label{manih1}
{\cal H}_1&=&-\hbar \Delta_a \int\!{\rm d}x\,\Psi_e^{\dag}\left(x\right)\Psi_e\left(x\right)-\hbar \Delta_c a^{\dag}a\\
&+&\left(\hbar \int\!{\rm d}x\,\left(g_0 \cos\left(kx\right)a+\Omega(x)\right)\Psi_e^{\dag}\left(x\right)\Psi_g\left(x\right)+{\rm H.c.}\right)\,,\nonumber
\end{eqnarray}
and which is here reported in the reference frame rotating at the laser frequency $\omega_p$. Here, $\Omega(x)$ denotes the strength of the coupling to the laser field (Rabi frequency), $g_0$ is the cavity-mode vacuum Rabi frequency, $\cos(kx)$ gives the spatial dependence of the cavity-mode function with wave vector $k$, while  $\Delta_{a}= \omega_{p}-\omega_{0}$ and  $ \Delta_{c}=\omega_p-\omega_c$ denote the detuning of the laser frequency from the atom and from the cavity-mode frequencies, respectively. The coherent dynamics of coupled atoms and electromagnetic field are then governed by the full  Hamiltonian $${\cal H}={\cal H}_0+{\cal H}_1\,.$$

This far we just considered Hamiltonian dynamics. We now introduce the sources of noise and dissipation, which in the present model are spontaneous radiative decay of the the electronic excited state $|e\rangle$ at the lifetime $\gamma$ and cavity losses at rate $2\kappa$ due to the finite transmittivity of the mirrors. Noise is here described within the Heisenberg-Langevin formalism, and the Heisenberg-Langevin equations for the atomic and field operators read~\cite{Larson_NJP08}
\begin{eqnarray}\label{pig}
& &\dot{\Psi}_g\left(x\right)=-\frac{{\rm i}}{\hbar}\left[\Psi_g\left(x\right),{\cal H}_0\right]\\
& &-{\rm i}\left(g_0 \cos\left(kx\right)a^{\dag}+\Omega(x)\right)\Psi_e\left(x\right)-\sqrt{\gamma}f_{\rm in }^\dag(t)\Psi_e\left(x\right)\nonumber \\\label{psie}
& &\dot{\Psi}_e\left(x\right)=-\frac{{\rm i}}{\hbar}\left[\Psi_e\left(x\right),{\cal H}_0\right]+\left({\rm i}\Delta_a -\frac{\gamma}{2}\right)\Psi_e\left(x\right)\,,\\
& &-{\rm i}\Psi_g(x)\left(g_0 \cos\left(kx\right)a+\Omega(x)\right)+\sqrt{\gamma}\Psi_g\left(x\right)f_{\rm in}(t)\,,\nonumber\\
& &\dot{a}=\left({\rm i} \Delta_c-\kappa\right)a -{\rm i} g_0 \int\!dx\,\cos\left(kx\right)\Psi_g^{\dag}\left(x\right)\Psi_e\left(x\right)\nonumber\\&&+\sqrt{2\kappa}a_{\rm in}(t)\,.\label{a}
\end{eqnarray}
Here, the noise operators $f_{\rm in}(t)$ and $a_{\rm in}(t)$ are the Langevin forces, with vanishing mean value and $\langle f_{\rm in}^{\dagger}(t)f_{\rm in}(t')\rangle=\langle a_{\rm in}^{\dagger}(t)a_{\rm in}(t')\rangle=0$, while $\langle f_{\rm in}(t)f_{\rm in}^{\dagger}(t')\rangle=\langle a_{\rm in}(t)a_{\rm in}^{\dagger}(t')\rangle=\delta(t-t')$, see for instance Ref.~\cite{carmichael}.

\subsection{Effective dynamics}

We now assume a time-scale separation, such that the electronic variables relax to a local steady state on a much shorter time scale than the one characterizing the dynamics of external and cavity degrees of freedom. This regime for the cavity degrees of freedom requires that the parameters satisfy $|\Delta_a+{\rm i}\gamma/2|\gg |\Omega|,g_0\sqrt{N},|\Delta_c|,\kappa$. Moreover, a change in the external degrees of freedom can be neglected over the typical time scale of the internal degrees of freedom provided that $|\Delta_a+{\rm i}\gamma/2|\gg \kappa_BT/\hbar$, with $\kappa_B$ Boltzmann's constant. In addition, we assume $|\Delta_a|\gg |\Omega(x)|,g_0\sqrt{N}\gg \gamma/2$, so that we can identify a time scale in which coherent evolution is taking place while dissipation due to spontaneous decay can be discarded. In these limits Eq.~(\ref{psie}) can be set to zero and at leading order gives
\begin{equation}
\label{psi:e:steady}
\Psi_e(x)\sim \Psi_g\left(x\right)\left[ag_0 \cos\left(kx\right)+\Omega(x)\right]/\Delta_a\,,
\end{equation}
such that Eq.~(\ref{a}) takes the form
\begin{eqnarray}
\dot{a}&=&-\kappa a+{\rm i}\left[\Delta_c-U_0\mathcal Y\right]a
-{\rm i}S_0\mathcal Z +\sqrt{2\kappa}a_{in}\,,\label{a:1}
\end{eqnarray}
where $U_0=g_0^2/\Delta_a$ is the maximal depth of the single-photon potential and $S_0=g_0\Omega/\Delta_a$ is the maximal amplitude of scattering a laser photon into the cavity mode by a single atom. In writing Eq.~(\ref{a:1}) we have assumed, moreover, that  $\Omega(x)=\Omega f(x)$, with $|f(x)|\le 1$. The spatial distribution of atoms and fields is now contained in the operators ${\cal Y}$ and ${\cal Z}$, which read
\begin{eqnarray}
\label{eq:Y}
&&{\cal Y}=\int {\rm d}x \cos^2(kx)\Psi_g^{\dagger}(x)\Psi_g(x)\,,\\
\label{eq:Z}
&&{\cal Z}=\int {\rm d}x f(x)\cos(kx)\Psi_g^{\dagger}(x)\Psi_g(x)\,,
\end{eqnarray}
and act on the Hilbert space of the atoms. The operators ${\cal Y}$ and ${\cal Z}$ are Hermitian and commute, $[{\cal Y},{\cal Z}]=0$, as they both depend solely on the atomic density. They are the quantum analogous of the semiclassical ``bunching parameter'' $\mathcal B=\sum_j\cos^2(kx_j)/N$ and of the ``spatial order parameter'' $\Theta=
\sum_j\cos(kx_j)/N$, respectively, which are defined for an ensemble of $N$ atoms at the positions $x_1,\ldots,x_N$, see for example~\cite{Asboth_PRA05}. These parameters characterize a semiclassical mean-field description of self-organization of the atoms in the cavity field, such that for $\Theta \to \pm 1$ the atoms are in a self-organized state at the even or odd antinodes of the cavity standing wave, while $\mathcal B$ provides their degree of localization about these points~\cite{Footnote}.

The cavity field can be eliminated from the matter-wave equations assuming a second characteristic time scale determined by the rate $|\Delta_c+{\rm i}\kappa|$, in which the cavity field approaches a local steady state. This requires that the effective coupling strength between matter waves and cavity field is smaller than this rate, namely $|\Delta_c+{\rm i}\kappa|\gg |S_0|\sqrt{N},|U_0|N$, and that the uncoupled matter wave dynamics is slower, $|\Delta_c+{\rm i}\kappa|\gg \kappa_BT/\hbar$. In this limit,  one finds $a\sim a^{(0)}$, with
\begin{equation}
\label{a:0}
a^{(0)}= \frac{S_0 {\cal Z}+{\rm i}\sqrt{2\kappa}a_{\rm in}(t)}{(\Delta_c-U_0{\cal Y})+{\rm i}\kappa}\,,
\end{equation}
where we have used that the operators in the numerator commute with the operators in the denominator.
Using in Eq.~(\ref{pig}) the solutions for $\Psi_e(x)$ and $a$, Eq.~(\ref{psi:e:steady}) and Eq.~(\ref{a:0}), at lowest order in perturbation theory we find
\begin{eqnarray}\label{psi:final}
&&\dot{\Psi}_g\left(x\right)=-\frac{{\rm i}}{\hbar}\left[\Psi_g\left(x\right),{\cal H}_0\right]-{\rm i}U_0\cos^2(kx) a^{(0)\dag}\Psi_ga^{(0)}\nonumber\\&&-{\rm i}S_0f(x)\cos(kx)a^{(0)\dag}\Psi_g-{\rm i}S_0f(x)\cos(kx)\Psi_ga^{(0)}\nonumber\\
&&-{\rm i}\frac{\Omega^2}{\Delta_a}f(x)^2\Psi_g\,.
\end{eqnarray}
Here $a^{(0)}$ is now a function of the atomic density and hence does not commute with the field operator $\Psi_g(x)$.

\subsection{Mechanical effects of the cavity field}

Equation~(\ref{a:0}) for the cavity-field operator $a^{(0)}$ gives the field state at leading order in a perturbative expansion, in which it is assumed that the cavity field follows adiabatically the matter-wave dynamics. Furthermore, it is seen that it is a function of the atomic density, which enters both in the numerator and in the denominator of the expression. In this work we will be interested in determining the atomic quantum ground state, which emerges from a coupled dynamics between cavity field and atoms, giving rise to the mechanical potential sustaining such state.

In order to gain further insight into the effect of the cavity potential on the atoms, we consider the semiclassical limit, which can be found from Eq.~(\ref{psi:final}). In this limit, we identify the terms which explicitly depend on the atomic position with mechanical forces. The corresponding potential takes the form
\begin{eqnarray}
\label{potential}
V(x)&=&V_1\cos^2(kx)+V_2\cos(kx)+\frac{\hbar \Omega^2 f(x)^2}{\Delta_a}\,,
\end{eqnarray}
with $V_1=\hbar U_0\langle a^{(0)\dagger}a^{(0)}\rangle$ and  $V_2=2\hbar S_0 {\rm Re}\{\langle a^{(0)}\rangle\}$. The third term on the right-hand-side of Eq.~(\ref{potential}) is the a.c.-Stark shift due to the laser field. In the following we assume that the external laser field drives the atoms uniformly and the Rabi frequency $\Omega$ does not depend on the position $x$ (i.e. $f(x)=1$). The Fourier transform of the potential is hence composed by a term at wavevector $2k$ whose amplitude is proportional to the number of photons, and which is due to the dispersive coupling between photons and atoms. This term corresponds to a potential with periodicity $\lambda/2$, analogous to an optical lattice in free space. Differing from an optical lattice in free space, the number of photons depends on the atomic density via the "bunching parameter". The second component is at wave vector $k$ and thus oscillates with periodicity $\lambda$. It originates from coherent Raman scattering of laser photons into the cavity field and its amplitude is hence proportional to the cavity electric field amplitude, but it also depends on the positions of the scatterers through the "spatial order parameter" $\Theta$. The two Fourier components give rise to an effective potential with spatial periodicity $\lambda$ as displayed in Fig.~\ref{Fig:1}. We remind that potential~(\ref{potential}) is found in a semiclassical limit. In this limit, the expression we find agrees with the ones derived in~\cite{Asboth_PRA05, Zippilli_APB04} within a semiclassical model.

The model of a semiclassical potential for the atoms does not take into account dissipative effects, which may be due for instance to cavity decay (while dissipation due to back-action of the resonator over the atoms is here {\it a priori} neglected as we assume the adiabatic limit). Indeed, in the dynamics of the atoms we find non-hermitian terms, which are characterized by the position-dependent coefficient $$\gamma'(x)={\rm Im}\{S_0a^{(0)}\}\cos(kx)\,,$$
scaling with the linewidth $\kappa$ of the resonator. Correspondingly, input noise gives rise to fluctuations in the height of the potential, which are described by the operator $a_{\rm in}(t)$ in Eq.~(\ref{a:0}). The effect of cavity quantum noise will be here discarded, assuming that the detuning $|\Delta_c|\gg\kappa$ (more precisely, $|\Delta_c|\gg\kappa,|U_0{\cal Y}|$, which is consistent with the assumption of adiabatic elimination of the cavity degrees of freedom). In this regime, the atoms move in a dispersive potential, which is produced by the dynamical Stark coupling with the cavity mode photons and by the Raman-scattered field of the laser.

\subsection{Tight-binding limit}

The motion and steady state of the semiclassical model discussed in the previous section has been studied in several theoretical works. Numerical simulations demonstrated that the system self-organizes, such that the atoms localize at the antinodes of the cavity field, where their coupling is maximum, according to an array with periodicity $\lambda$~\cite{Domokos_PRL02}. This behavior appears when the laser intensity exceeds a critical value~\cite{Domokos_PRL02, Asboth_PRA05}, and has been confirmed experimentally~\cite{Black_03}. Localization of the atoms may take place around two possible set of minima, one centered at the positions $x_{2j}^{(0)}=j\lambda$ (where $\cos(kx_{2j}^{(0)})=+1$), with $j$ integer, and the other at the positions $x_{2j+1}^{(0)}=(2j+1)\lambda/2$ (where $\cos(kx_{2j+1}^{(0)})=-1$), while the cavity field is a coherent state with amplitude proportional to the pump~\cite{Zippilli_PRL04,Zippilli_EPJD04}. In the limit in which the atoms are well localized at the minima one can expand potential~(\ref{potential}) till second order in the fluctuations of the particles position. For a particle localized at $x_j^{(0)}$ the potential takes the form
\begin{equation}
V(x)\simeq V^{(0)}+V^{(2)}x^2
\end{equation}
where $x=x_j-x_j^{(0)}$ and $V^{(0)}=V(x_j^{(0)})$, while
\begin{equation}
\label{potential:sc}
V^{(2)}=-\frac{\hbar k^2 S_0^2\langle{\cal Z}\rangle}{(\Delta_c-U_0\langle{\cal Y}\rangle)^2+\kappa^2}\left(U_0\langle{\cal Z}\rangle+(-1)^j(\Delta_c-U_0\langle{\cal Y}\rangle)\right)
\end{equation}
where the potential due to the other particles enters in a mean-field approach through the mean value of operators $\cal Z$ and $\cal Y$. Note that when the atoms are self-organized at the set of positions $\{x_{2j}^{(0)}\}$ (respectively, $\{x_{2j+1}^{(0)}\}$), then the sign of $\langle\cal Z\rangle$ is positive (respectively, negative).

Consistently with the approximations made so far, the dynamics can be characterized by the dispersive dipolar potential provided that $|U_0\langle{\cal Y}\rangle|,|U_0\langle{\cal Z}\rangle|\ll |\Delta_c|$, which allows for the adiabatic elimination of the cavity field from the matter-wave equations. This implies that, for the regime we consider, the sign of this term of the potential is solely determined by the sign of the detuning between cavity and pump. One hence finds that the position $x_j^{(0)}$ are minima provided that $\Delta_c<0$. This behavior can be understood in terms of positive feedback of the system, which is warranted whenever the sign of $\langle {\cal Z}\rangle$ is opposite to the one of the harmonic potential $V^{(2)}$, and hence when $\Delta_c<0$. This latter condition leads to the property that the conservative force, due to the potential, is attractive~\cite{Zippilli_APB04}. It is a sufficient condition, provided that the detuning $\Delta_a$ between atom and laser is negative, $\Delta_a<0$ (and consequently $U_0$ is negative), as it is visible from the numerator of Eq.~(\ref{potential:sc}) (note that for the situations here considered, $|\langle {\cal Z}\rangle| > \langle {\cal Y}\rangle$, see also Ref.~\cite{Domokos_EPJD08}). Indeed, in this regime the minima of the potential, formed by light scattering of the atoms, are also minima of the standing-wave potential of the cavity field, so that the attractive forces of both contributions add up to confine the atoms. On the contrary, when $\Delta_a>0$ the opposite situation is realized: the minima of the potential due to light scattering are now maxima of the cavity standing-wave potential: the term proportional to $U_0$ in the numerator of Eq.~(\ref{potential:sc}) has opposite sign than the detuning $\Delta_c$, and the overall effect is to make the effective potential, resulting from the two contributions, shallower. An extensive discussion on the semiclassical forces as a function of the detunings can be found in Refs.~\cite{Zippilli_APB04, Asboth_PRA05}.

Finally, the amplitude of the semiclassical detuning depends nonlinearly on the atomic density through  quantity $\Delta_c-U_0{\cal Y}$ in the denominator. Bistability effects due to this nonlinearity have been discussed for the situation, in which the cavity is pumped by a laser~\cite{Larson_PRL08, Larson_PRA08,Meystre_PRA09,Stamper-Kurn_PRL07,Esslinger_Science08,Ritter_APB09}. In this article we will focus on the regime where $|\Delta_c-U_0\langle{\cal Y}\rangle|\gg\kappa$, when the potential is  conservative and dissipative effects can be neglected. This parameter regime is far away from the bistability region. Correspondingly, this term will be treated as a small correction of the semiclassical potential in the framework of perturbation theory.

\section{Effective dynamics in the cavity potential}
\label{Sec:Bose-Hubbard}

\subsection{Derivation of the effective Hamiltonian}

We now assume that the atoms are tightly confined in one set of minima of potential~(\ref{potential}), say, at the position $x_{2j}=j\lambda$. This assumption is clearly based on the semiclassical dynamics, as a potential cannot be simply singled out from Eq.~(\ref{psi:final}). Our aim is to identify the quantum ground state by considering the full quantum dynamics. For this purpose, starting from the assumption of tight confinement, we perform a Wannier decomposition of the atomic field operator,
\begin{equation}\label{wannier}
\Psi_g\left(x\right)=\sum_{j} w_j(x)b_{j}\,,
\end{equation}
where operator $b_j$ annihilates a particle at the site centered in $x_{2j}^{(0)}$, and $w_j(x)$ is the Wannier function, taken to be real valued and with center in $x_{2j}^{(0)}$. This decomposition is based on the assumption that the atoms are in the lowest band of the semiclassical potential, which is justified at ultralow temperatures. Within the decomposition of Eq.~(\ref{wannier}), Hamiltonian~(\ref{H:0}) takes the form ${\cal H}_0\simeq{\cal H}_{0}^{BH}$,  with
\begin{eqnarray}\label{no}
{\cal H}_{0}^{BH}=E_{0}N+E_{1}B+\frac{1}{2}U\sum_i n_{i}\left(n_{i}-1\right)-\mu_0 N\,,
\end{eqnarray} where $N=\sum_{j}b^{\dag}_{j}b_{j}$ is the atom number operator, and  $B=\sum_{j}\left(b^{\dag}_{j}b_{j+1}+{\rm H.c.}\right)$ is the hopping operator, while the coefficients read
\begin{eqnarray}\label{cm}
E_{\ell}&=&\int\!dx\, w_i\left(x\right) \left(-\frac{\hbar^{2}}{2m}\frac{\partial^2}{\partial x^2}\right) w_{i+\ell} \left(x\right)\\
U&=&g_{1D}\int\!dx\, w_i(x)^4\,,\label{u:1}
\end{eqnarray}
and $g_{1D}=g_g$ is the scattering strength in one dimension. In writing Eq.~(\ref{no}) we made the nearest-neighbor approximation, and introduced the chemical potential $\mu$, assuming a grand-canonical ensemble. We remark that the Bose-Hubbard expansion of Hamiltonian~(\ref{no}) relies on localization of the atoms at the minima of the potential inside the cavity.

We now use Eq.~(\ref{wannier}) in  Eq.~(\ref{psi:final}), and determine the equations of motion for the operators $b_l$ by multiplying with $w_l(x)$ and integrating over the position. We find
\begin{eqnarray}\label{eq:b}
\dot{b}_l&=&-\frac{{\rm i}}{\hbar}\left[b_l,{\cal H}_0^{BH}\right]-{\rm i}U_0{\cal A}^{(0)\dag}\left[J_{0}b_l+J_{1}{\cal B}_{l}\right]{\cal A}^{(0)}\nonumber\\
&-&{\rm i}S_0{\cal A}^{(0)\dag}\left[Z_{0}b_l+Z_{1}{\cal B}_{l}\right]-{\rm i}S_0\left[Z_{0}b_l+Z_{1}{\cal B}_{l}\right]{\cal A}^{(0)}\nonumber\\
&-&{\rm i}\frac{\Omega^2}{\Delta_a}b_l\,,
\end{eqnarray}
which is written in the regime in which $|\Delta_c|\gg |U_0{\cal Y}|$. Here, ${\cal B}_l=b_{l+1}+b_{l-1}=[b_l,B]$, and ${\cal A}^{(0)}\simeq a^{(0)}$, such that
\begin{equation}
{\cal A}^{(0)}=\frac{S_0(Z_0N+Z_1B)}{\Delta_c+{\rm i}\kappa}\left(1+\frac{U_0}{\Delta_c+{\rm i}\kappa}(J_0N+J_1B)\right)\,,
\end{equation}
which considers only nearest-neighbor coupling and where we used
\begin{eqnarray}
{\cal Z}&\simeq&Z_0 N+ Z_1 B\,,\\
{\cal Y}&\simeq&J_0 N+J_1 B\,,
\end{eqnarray}
with
\begin{eqnarray}\label{me}
Z_{\ell}&=&\int\!{\rm d}x\, w_i(x) \cos \left(kx\right) w_{i+\ell}(x)\,,\\
J_{\ell}&=&\int\!{\rm d}x\, w_i(x) \cos^{2}\left(kx\right) w_{i+\ell}(x)\,,
\end{eqnarray}
and $\ell=0,1$. At first order in the expansion in the small parameter $U_0/|\Delta_c|$, Eq.~(\ref{eq:b}) can be exactly cast in the form
\begin{equation}
\dot{b}_l=-\frac{{\rm i}}{\hbar}\left[b_l,{\cal H}_0^{BH}\right]-\frac{{\rm i}}{\hbar}\left[b_l,{\cal H}_{\rm CQED}^{BH}\right]\,,
\end{equation}
with
\begin{eqnarray}
\label{H:CQED:1}
& &H_{\rm CQED}^{BH}=\frac{\hbar S_0^2}{\Delta_c}\left[(Z_0N+Z_1B)^2+\frac{U_0}{\Delta_c}Z_0\right.\\
& &\left.\times\left(Z_0J_0N^3+Z_0J_1NBN+J_0Z_1(N^2B+BN^2)+{\rm o}(Z_1^2)\right)\right]\,,\nonumber
\end{eqnarray}
where we consider $\kappa\ll|\Delta_c|$ and we truncated the second term in an expansion at first order in $|Z_1U_0N/\Delta_c|$ and  $|J_1U_0N/\Delta_c|$. Note that in expression~(\ref{H:CQED:1}) the operators are symmetrically ordered, so to obtain Eq.~(\ref{eq:b}).

The total dynamics of the system is now rendered by the Hamiltonian
\begin{equation}
\label{H:eff}
H_{\rm eff}=H_{0}^{BH}+H_{\rm CQED}^{BH}\,,
\end{equation}
where the first term of the right-hand-side, defined in Eq.~(\ref{no}), describes the hopping between sites, due to the quantum fluctuations, and the on-site interaction emerging from $s$-wave scattering, while the second term contains the contributions due to the mechanical effects of the laser and cavity potential, which determine and sustain the atomic pattern. We note that Hamiltonian~(\ref{H:CQED:1}) is similar to the Bose-Hubbard Hamiltonian, however, its terms are nonlinear in the number and hopping operators. Moreover, it differs also from the usual Bose-Hubbard model due to the appearance of a term proportional to the squared of the hopping operator, $B^2$.

The nonlinearity in Eq.~(\ref{H:CQED:1}) is due to the coupling with the cavity field, which scales with the number of atoms and which determines the confining potential: the larger is the atomic number the stronger is the coupling. This effect is exquisitely due to cavity quantum electrodynamics, and originates from the constructive interference with which the atoms scatter the laser photons into the cavity mode. We also note that, when considering higher order terms in $U_0/\Delta_c$, an exact effective Hamiltonian -which is a function solely of operators $N$ and $B$ fulfills Eq.~(\ref{eq:b})- cannot be found unless one resorts to a suitable thermodynamic limit~\cite{Larson_NJP08}. In fact, these terms emerge from the nonlinear coupling between photons and atoms inside the resonator, in which the atoms act as a refractive index. The effect is hence highly nonlocal, and it is significant when the strong coupling regime is warranted. An immediate consequence is that Wannier functions will depend on the particle number. Most remarkable is the fact that the Hamiltonian is quadratic in the hopping operator, even when the nonlinearity deriving from the atomic refractive index, $U_0{\cal Y}$, is negligible. This property is due to the collective scattering of the atoms into the resonator. Nevertheless, it does not affect substantially the property of the ground state at large potential depth (small tunneling rates) as we will show.

\subsection{Mapping to a Bose-Hubbard model}

For the purpose of studying the quantum phases of the many-body system, we define a thermodynamic limit. This is here so defined, that the density of the atoms is kept fixed as the particle number $N$ and the cavity-mode volume go to infinity. Denoting by $2K$ the total number of sites of the lattice, such that in the considered configuration there are $K$ lattice cells, the cavity coupling strength is inversely proportional to the squared root of the cavity mode volume, and thus scales as $g_0\propto1/\sqrt{K}$. The other parameters in Eq.~(\ref{H:eff}) scale as $U_0=u_0/K$ and $S_0=s_0/\sqrt{K}$, where $u_0$ and $s_0$ are constants, while the number operator is now $N=K n_0$, with $n_0$ the density operator, giving the number of atoms per site. This scaling is such that, as $K\to \infty$, the parameters in Hamiltonian~(\ref{H:eff}) depend only on the atomic density. We also note that in this thermodynamic limit a Bose-Hubbard form can be also derived when the terms scaling with $U_0$ are taken into account, see for instance~\cite{Larson_NJP08} for a discussion. In this thermodynamic limit we rewrite Hamiltonian~(\ref{H:eff}) in the Bose-Hubbard-like form,
\begin{equation}
H=-t_1[n_0]B+t_2[n_0]B^2+\frac{U[n_0]}{2}\sum_in_i(n_i-1)-\mu[n_0]N\,,
\end{equation}
where
\begin{eqnarray}
&&t_1[n_0]=-E_1-n_0\hbar\frac{s_0^2}{\Delta_c}\left[2Z_0Z_1+n_0\frac{u_0}{\Delta_c}(Z_0^2J_1+2J_0Z_0Z_1)\right]\,,\nonumber\\
\label{t1}\\
&&t_2[n_0]=\frac{\hbar s_0^2}{\Delta_c}Z_1^2\left[1+{\rm O}(u_0/\Delta_c)\right]\,,\label{t2}\\
&&\mu[n_0]=\mu_0-E_0-n_0\hbar\frac{s_0^2}{\Delta_c}Z_0^2\left(1+n_0\frac{u_0}{\Delta_c}J_0\right)\,,\label{mu}
\end{eqnarray}
and where $U$, $Z_j$, and $J_j$ are functions of $n_0$ through the Wannier functions. Insight into the dependence of the coefficients on the density can be gained using a (modified) Gaussian ansatz in place of the Wannier functions, as reported in Appendix~\ref{App:A}.

Differing from the ordinary Bose-Hubbard model~\cite{Fisher_PRB89}, in the case of self-organized atomic patterns the coefficients entering the effective BH Hamiltonian do depend on the atomic density. This property implies that the atomic density is not solely determined by the chemical potential~\cite{Larson_PRL08,Larson_NJP08}.
We also observe that, in addition to the nearest-neighbor hopping arising from the kinetic energy, there is a hopping term due to the cavity potential which contains a term proportional to $B^2$. This latter contribution describes long-range correlations among the atoms mediated by the scattered photons. Close inspection shows that, in the tight-binding regime, its coefficient is of higher order with respect to the coefficient multiplying $B$, see App.~\ref{App:A}. Within the considered thermodynamic limit, this term gives rise to a small correction to the spectrum of the lowest excitations and corresponding eigenstates of the Mott-insulator states, which must, however, be taken into account when determining the ground state of the system for larger values of the coefficient $Z_1$, scaling the tunneling coefficient $t_2$ and some terms of $t_1$.

\section{Incompressible phase of self-organized atomic patterns}\label{Results}

We now study whether the self-organized atomic pattern exhibits an incompressible ground state. Therefore, we focus onto the regime in which self-organization has set in, namely, when the laser amplitude exceeds the threshold value~\cite{Asboth_PRA05, Domokos_EPJD08} and the laser frequency is smaller than the cavity-mode frequency, $\Delta_c<0$. The atoms are assumed to be localized at the set of minima $\{x_{2j}=j\lambda\}$, such that $Z_0>0$.

In order to display the phase diagram for this system, we consider the chemical potential and the amplitude of the pump field $s_0$, which indirectly determines the tunneling coefficients. In fact, for $s_0\to\infty$ the intracavity field potential is deepest and tightly localizes the atoms at their minima, and the tunneling parameters $t_1$ and $t_2$, Eqs.~(\ref{t1}) and~(\ref{t2}), vanish correspondingly (as one can check by considering the dependence of the coefficients $Z_j$ and $J_j$ on $s_0$, see Appendix~\ref{App:A}).

We apply the strong coupling expansion developed in Ref.~\cite{Monien_94} in order to determine the size of the incompressible states as a function of the pump amplitude and of the chemical potential. For this purpose we start by considering the limit $s_0\to\infty$, assuming the system to be in a Mott-insulator state with an integer occupation per site $n_0$. We calculate the free energy of the Mott-insulator state $E_M(n_0)$ and of the defect states $E_\pm(n_0)$, which are obtained by adding or removing a particle to the Mott-insulator state. The energy of the defect state is found by means of degenerate perturbation theory, expanding in power of the small parameter $t_1/U$ up to third order. In the parameter regime we consider, this expansion corresponds also to a first-order expansion in the parameter $t_2$. The phase boundaries between the Mott-insulator and the compressible phase correspond to the situation where the energy difference between the defect state and the Mott-insulator state vanishes, namely, when
\begin{equation}
E_\pm(n_0)- E_M(n_0)=0\,.
\end{equation}
The values of the chemical potentials $\mu^+_{(n_0)}$ and $\mu^-_{(n_0)}$, at which the Mott-insulator phase with integer occupation $n_0$ becomes unstable when adding and removing a particle, respectively, fulfill the equations
\begin{eqnarray}\label{mup}
E_+(n_0)-E_M(n_0)&=&-\mu^+_{(n_0)}+U n_0-2t_1(n_0+1)+\frac{t^2_1}{U }n_0^2\nonumber\\
&+&\frac{t^3_1}{U ^2}n_0(n_0+1)(n_0+2) \nonumber\\
&+&t_2 2(2n_0+1)(n_0+2)=0\,,\\
\label{mum}
E_-(n_0)-E_M(n_0)&=&\mu^-_{(n_0)}-U (n_0-1)-2t_1n_0\nonumber\\&+&\frac{t^2_1}{U }(n_0+1)^2
+\frac{t^3_1}{U ^2}n_0(n_0+1)(n_0-1) \nonumber\\&+&t_2 2(2n_0+1)(n_0-1)=0\,.
\end{eqnarray}
When $\mu^{+}_{(n_0)}>\mu^{-}_{(n_0)}$, then $\mu^+_{(n_0)}$ and $\mu^-_{(n_0)}$ determine the upper and lower boundaries of the parameter region where the system is in the Mott-insulator state. The phase diagrams reported afterwards are computed using the analytical formulae of the strong coupling expansion, Eqs.~(\ref{mup}) and~(\ref{mum}) in the thermodynamic limit, fixing the values of the parameters $u_0$ and $s_0$. The coefficients are evaluated using a modified Gaussian ansatz in place of the Wannier functions, see Appendix for detail.

Figure~\ref{fig:2} displays the phase diagram for the considered system, showing three incompressible, Mott-insulator phase regions with $n_0=1,2,3$, for $\Delta_c=-50\kappa$ and $u_0=-0.1\kappa$. When plotted as a function of $1/s_0$, the Mott-insulator regions become larger as a function of the laser amplitude as the number of atoms increases. This result can be understood with simple arguments, in fact the trapping potential is constituted of the light coherently scattered by the atoms: As the number of atoms increases, the cavity-field intensity, for a given pump amplitude, is larger, thus the potential becomes deeper and the configuration more stable. Another remarkable property is the appearance of gaps between the Mott-insulator lobes. In the limit of vanishing tunneling coefficients $t_{1,2}\to 0$ (corresponding to $s_0\to \infty$), this behavior implies that the degeneracy among states with different densities, found in the ordinary BH model, is here removed. This behaviour results from the dependence of the single-particle, on-site energy on the average density. In particular, while inside the lobes the density per site $n_0$ takes integer values, in the gap between the lobes the density changes continuously in the interval $r<n_0<r+1$ (with $r=1,2,3,\dots$ integer number), as shown in Fig.~\ref{fig:3}. The region outside the Mott-insulator lobes is presumably a peculiar superfluid state, which entails photonic excitations. A study of its properties close to the self-organization threshold has been presented in Refs.~\cite{Domokos_EPJD08,Domokos_PRL02}. The appearance of the gap is understood, when one considers the dependence of the coefficients on the density. In fact, at $s_0\to \infty$ (corresponding to putting the tunneling elements to zero), the Mott-insulator region with $n_0=1$ is found for values of $\mu$ such that $0<\mu_1<U_1$, where $\mu_1=\mu[n_0=1]$ and $U_1=U[n_0=1]$ (see Eq.~(\ref{u:1})). Similarly, the region at $n_0=2$ is found for $U_2<\mu_2<2U_2$ (with $U_2=U[n_0=2]$). The relation $U_1<U_2$ ($U_j<U_{j+1}$) leads to the appearance of parameter regions where the system is compressible even at zero tunneling. This relation is a consequence of the fact that, for a fixed amplitude of the laser field, the intracavity field becomes larger as the number of atoms (scatterers) is increased.

Figure~\ref{fig:4} displays different phase diagrams, obtained for various values and signs of the detuning $\Delta_a$, keeping $\Delta_c$ fixed. We first focus on the case in which $\Delta_a<0$, when the potential due to the dynamical a.c.-Stark shift, induced by the coupling with the cavity mode, adds up to the potential due to light scattering, giving rise to a stronger attractive dipole force. In this case, we observe that for larger values of $u_0$ the gap between the chemical potentials of Mott-insulator states at different densities increases. Moreover, the larger is $u_0$, the larger is the area covered by Mott-insulator states at larger densities: incompressible states are found at smaller values of the laser amplitude. The opposite tendency is found when $\Delta_a>0$, and hence when the potential due to the a.c.-Stark shift competes with the scattering potential. In this case, as $|u_0|$ is increased, the gap between the Mott-insulator lobes becomes smaller, while the area covered by Mott-insulator states with larger densities decreases. In particular, incompressible states are observed at larger values of the laser pump amplitude. This behavior can be understood from the dependence of the tunneling coefficient, Eq.~(\ref{t1}), on the various terms: The term proportional to $u_0$ scales with the onsite density, and it hence becomes more important as $n_0$ is increased. For values of the detuning, such that the cavity potential adds up to the scattering potential ($\Delta_a<0$), at a fixed laser intensity the confinement is larger for a larger density of atoms. Correspondingly, the parameter region where the system is compressible becomes larger. On the contrary, for $\Delta_a>0$ the cavity potential tends to make the scattering potential shallower, the confinement tends to depend less strongly on the density and the gap between the Mott-insulator lobes is smaller.

\begin{figure}[htbp]
\begin{center}
\includegraphics[scale=0.8]{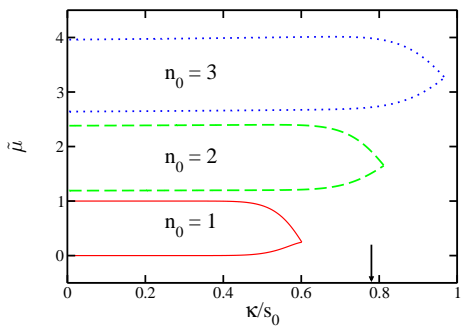}
\caption{(color online) Phase diagram, showing the Mott-insulator states at different densities $n_0=1,2,3$ in the plane reporting the chemical potential $\tilde{\mu}=\mu/U_1$, Eq.~(\ref{mu}), and the inverse of the laser amplitude, $1/s_0$ (in units of cavity loss rate $\kappa$). The parameters are $\Delta_c=-50\kappa$ and $u_0=-0.1\kappa$. The arrow indicates the threshold value of the laser amplitude, for which the system self-organizes~\cite{Domokos_EPJD08}. The chemical potential is here reported in units of the on-site energy $U_1$, corresponding to the value of Eq.~(\ref{u:1}) for the density of one atom per site, with $g_{1D}/(E_R\lambda)=3.74\cdot10^{-4}$ and $E_R$ the recoil energy, considering a gas of ${}^{87}\!Rb$ atoms whose dipole-transition, at wavelength $\lambda=830$ nm, is coupled to the cavity mode.}
\label{fig:2}
\end{center}
\end{figure}

 \begin{figure}[htbp]
\begin{center}
\includegraphics[scale=0.8]{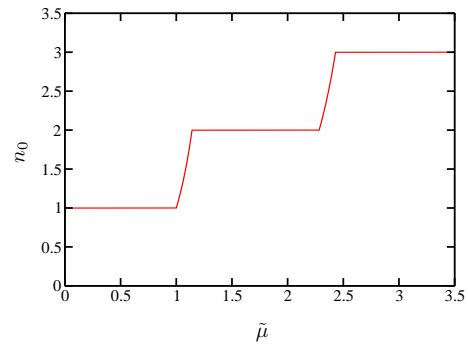}
\caption{(color online) On-site density $n_0$ as a function of the chemical potential $\tilde{\mu}$ (in units of the on-site energy $U_1$) for $s_0\to \infty$ (corresponding to $t_1,t_2\to 0$). Same parameters as in Fig.~\ref{fig:2}.}
\label{fig:3}
\end{center}
\end{figure}


\begin{figure}[htbp]
\begin{center}
\includegraphics[scale=0.8]{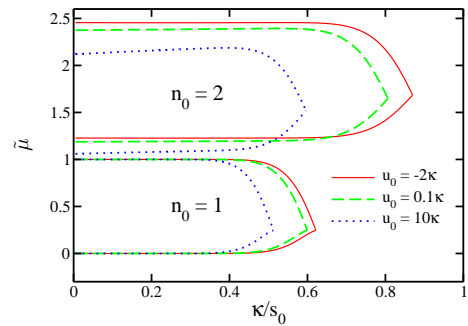}
\caption{(color online) Mott-insulator states at $n_0=1,2$ in the plane $\tilde{\mu}/U_1$-$\kappa/s_0$ for $\Delta_c=-50\kappa$ and (i) $u_0=-2\kappa$ (red solid line), (ii) $u_0=0.1\kappa$ (green dashed line), and (iii) $u_0=10\kappa$ (blue dotted line). The other parameters are the same as in Fig.~\ref{fig:2}.
}
\label{fig:4}
\end{center}
\end{figure}

We remark that, while in this model atom-atom interaction is essentially $s$-wave scattering, it should be considered that, when the atomic occupation of each site exceeds unity, then dipole-dipole interaction, including superfluorescence, will be relevant. In our model we neglect collective effects in the spontaneous decay outside of the cavity. In particular, for the parameter regime we consider, incoherent processes are suppressed for the large atomic detunings we choose. On the other hand, one can identify parameter regimes where coherent coupling between the atoms at a site may be relevant, and which should be included in an effective onsite interaction term.

We finally estimate the parameter regime required for observing our predictions, considering the setup reported in Ref.~\cite{Ritter_APB09}. Here, ultracold $^{87}$Rb atoms are loaded inside a resonator, where a mode couples quasi-resonantly with the transition at wave length $\lambda=780$ nm. The parameters are $g_0/\pi=14.1$ MHz, $\kappa/2\pi=1.3$ MHz and $\gamma/2\pi=3$ MHz. Taking $|\Delta_a|/2\pi=60$ GHz, $\Omega/2\pi=60$ MHz, we find $U_0/2\pi\sim 0.03$ MHz and $S_0/2\pi=0.1$ MHz, while the effective rate of spontaneous emission is of the order of few KHz. For a number of atoms $N\sim 10^4$, the relation $S_0\sqrt{N}\gg\kappa$ is fulfilled, warranting the existence of a time-scale in which the atoms may experience a conservative potential formed by their scattered photons. The relation $|\Delta_c|\gg |U_0|N,\kappa$ is satisfied taking $|\Delta_c|/2\pi=100$ MHz. For these parameters, the value $s_0/\kappa=10$ at which incompressible phases are found in Figs.~\ref{fig:2} and~\ref{fig:4}, corresponds to $S_0\sqrt{N}\sim 10\kappa$. This latter value corresponds to the one found for the experimental parameters in~\cite{Ritter_APB09}, showing that the regime, where quantum effects in self-organized atomic patterns are visible, can be accessed with existing experimental setups.

\section{Conclusions}\label{Conclusions}

The quantum ground state of a self-organized atomic pattern, confined by the potential of the field scattered into the cavity mode, can be incompressible provided that the laser intensity exceeds a certain critical value, which is found above the threshold value at which the system self-organizes. In addition, intervals of values of the chemical potential are identified, in which the state is compressible even when the tunneling vanishes. These appears as gaps between the Mott-insulator lobes in a phase diagram we derive, and their origin can be understood from the competition between confinement, which is steeper at large atomic density for a given laser intensity, and interparticle collisions, whose effective strength also depends on the density via the localization of the atomic wave function in the intra-cavity field potential. Our analysis is based on an effective Bose-Hubbard model, according to the assumption that the atoms are tightly bound at the minima of the intra-cavity field potential. Since the height of such potential depends on the atomic density, the coefficients of the model depend on the atomic density giving rise to the peculiarities encountered in the phase diagram. Comparison with typical parameters of experimental setups, dealing with ultracold atoms inside of optical cavities, shows that the prediction here made could be experimentally tested in existing setups.

Detection of the quantum properties of the system may be performed by measuring the light using a probe field~\cite{Mekhov_NP07,Mekhov_PRL07,Esslinger_09}, in a direction of emission corresponding to a Bragg scattering angle. A spectral analysis of the probe intensity will provide information on the state of the system, which can be studied as a function of the laser pump intensity, and hence of the tunneling rate~\cite{Rey_PRA05,Rist_PRA10}. Sweeping the phase diagram in the direction $\tilde{\mu}$ may be performed in various ways. One possibility is to implement a shift of the atomic ground state by means of a magnetic field or by off-resonant coupling with a third electronic state.

In this work we focused on the quantum ground state of the self-organized structure, arising by driving the atoms with a strong, transversal laser field. The mechanical potential results from the scattering potential, with periodicity equal to the wavelength $\lambda$, and from the potential due to the cavity field mode, with periodicity equal to the wavelength $\lambda/2$. The regime we considered is the one in which the scattering potential is the strongest. In this regime, it is expected that the light at the cavity output is coherent, while the intensity is independent on the number of atoms when the strong-coupling regime is warranted~\cite{Zippilli_PRL04,Zippilli_EPJD04}. An interesting question is what are the properties of the emitted light when the cavity-field mode potential dominates. In this regime, in a semiclassical model it has been found that the light at the cavity output can be squeezed, and in general anti-bunched~\cite{Fernandez_PRA07,Habibian}. An intriguing question is whether and how these properties are modified, when the atomic quantum statistics is relevant to the dynamics of atom-photon interactions.

\subsection*{Acknowledgments}

We are grateful to Maciej Lewenstein and Helmut Ritsch for stimulating discussions and helpful comments. We also acknowledge discussions and comments by Peter Domokos, Christoph Maschler, Igor Mekhov, Wolfgang Niedenzu, Duncan O'Dell, and Andras Vukics. Support by the ESF (EUROQUAM "CMMC"), the European Commission (EMALI, MRTN-CT-2006-035369; SCALA, Contract No.\ 015714) and the Spanish Ministerio de Educaci\'on y Ciencia (Consolider Ingenio 2010 QOIT; QNLP, FIS2007-66944; Acci\'on Integrada HU2007-0013; Ramon-y-Cajal and Juan-de-la-Cierva programs), and the Swedish government are acknowledged. GM acknowledges support by the German Research Council (Heisenberg professorship).

\begin{appendix}

\subsection{Coefficients in the Gaussian Approximation}
\label{App:A}

The solution of the Bose-Hubbard model requires the evaluation of the coefficients, which are integrals containing Wannier functions. The Wannier functions are solutions of the Schr\"odinger equation for a single particle in presence of the potential in Eq.~(\ref{potential}), which itself depends on the spatial distribution of the atoms inside the cavity. The problem is hence nonlinear. We solve it by combining an analytical and a numerical approach. In first instance, we replace the Wannier functions by modified Gaussian functions, $w_i(x)\to \tilde{w}_i(x)$, with
\begin{eqnarray}
\label{Wannier:modify}
\tilde{w}_i(x)=\left(1+\frac{3\delta^2}{4}\right)g_i(x)-\frac{\delta}{2}(1+\frac 32\delta^2)\left(g_{i+1}(x)+g_{i-i}(x)\right)\,,\nonumber\\
\end{eqnarray}
and
\begin{equation}
g_i(x)=\frac{1}{\sqrt[4]{\pi \sigma^2}}e^{-\frac{(x-x_i)^2}{2\sigma^2}}
\end{equation}
is a normalized Gaussian function of width $\sigma$. The form of the modified Gaussian function in Eq.~(\ref{Wannier:modify}) warrants orthonormality up to second order in $\delta$, where $\delta=\int {\rm d}x g_i(x)g_{i+1}(x)={\rm e}^{-\pi^2/(k^2\sigma^2)}$ and where the overlap between second neighbors is of the order ${\rm O}(\delta^4)$. We remark that the coefficient in~(\ref{Wannier:modify}) at second order in $\delta$ warrants the correct normalization. The width $\sigma$ of the Gaussian function $g_i(x)$ is found by minimizing the single-particle energy $E$, for the Schr\"odinger equation in presence of potential~(\ref{potential}). For later convenience we introduce the dimensionless parameter
\begin{equation}
\label{Eq:y}
y=k^2\sigma^2\,,
\end{equation}
which is $y\ll1$, consistently with our assumptions.
In terms of the parameter $y$ this consists in finding the minimum of the equation
\begin{equation}
E(y)=\frac{E_R}{2 y}+F(y)-\frac{G(y)}{2}y\,,
\end{equation}
where $E_R=\hbar^2 k^2/2m$ is the recoil energy and we introduced the functions
\begin{eqnarray}
&&F(y)=\tilde{V}_1+\tilde{V}_2\,,\\
&&G(y)=\tilde{V}_1+\tilde{V}_2/2\,,
\end{eqnarray}
where $\tilde{V}_1$ and $\tilde{V}_2$ are
\begin{eqnarray}
&&\tilde{V}_1=\frac{\hbar U_0S_0^2{\cal Z}(y)^2}{\kappa^2+(\Delta_c-U_0{\cal Y}(y))^2}\,,\\
&&\tilde{V}_2=\frac{2\hbar S_0^2{\cal Z}(y)(\Delta_c-U_0{\cal Y}(y))}{\kappa^2+(\Delta_c-U_0{\cal Y}(y))^2}\,.
\end{eqnarray}
Note that the $\cos$-functions in Eq.~(\ref{potential}) have been expanded up to second order in the displacement from the minima. Minimization with respect to $y$ leads to the nonlinear equation
\begin{equation}
y^2=\frac{E_R}{2F^{\prime}(y)-yG^{\prime}(y)-G(y)}\,,
\end{equation}
with $F^{\prime}(y)=dF(y)/dy$ and $G^{\prime}(y)=dG(y)/dy$, and which is solved numerically by recursion.

Using the value which minimizes the single-particle energy, $y:=y_{\rm min}$, we evaluate the coefficients we need in order to study the problem. They are reported below as a function of $y$:
\begin{eqnarray}
&&E_0=\frac{E_R}{2y}\left(1+\delta^2\frac{4\pi^2}{y}\right)\,,\\
&&J_0=\frac{1+{\rm e}^{-y}}{2}\,,\\
&&Z_0=(1+4\delta^2){\rm e}^{-y/4}\,,\\
&&E_1=-\frac{E_R \delta \pi^2}{y^2}\left(1+\frac94 \delta^2\right)\,,\\
&&J_1=0\,,\\
&&Z_1=-2\delta{\rm e}^{-y/4}\left(1+\frac94 \delta^2\right)\,,\\
&&U=g_{1D}k\frac{1+3\delta^2-4\delta^{\frac{5}{2}}}{\sqrt{2\pi y}}\,,
\end{eqnarray}
where $E_R=\hbar^2 k^2/2m$ is the recoil energy and we discarded terms of higher order than $\delta^2$. Note that the coefficients $Z_j$ are here reported for the pattern localized at the positions $x_{2j}^{(0)} =j\lambda$.
\end{appendix}


\begin{thebibliography}{100}
\bibitem{Footnote}
A uniform distribution of atoms along the cavity axis corresponds instead to $\mathcal B\sim 1/2$, and correspondingly $\Theta\approx 0$. On the other hand, one could also have the situation in which $\Theta\approx 0$ while $\mathcal B\ll 1$, corresponding to the atoms localized at both even and odd antinodes of the cavity standing wave potential.
\end{thebibliography}
\end{document}